\documentclass[genTeX]{nrc1}

\newcommand{\nco}{\newcommand}
\nco{\beq}{\begin{equation}} \nco{\eeq}{\end{equation}}
\nco{\beqa}{\begin{eqnarray}} \nco{\eeqa}{\end{eqnarray}}
\nco{\lsim}{\mbox{\raisebox{-.6ex}{~$\stackrel{<}{\sim}$~}}}
\nco{\gsim}{\mbox{\raisebox{-.6ex}{~$\stackrel{>}{\sim}$~}}}

\def\exd{{\rm d}}

\volyear{83}{2005}
\journal{Can. J. Phys.}
\received{June 5, 2005}
\accepted{?????}
\begin{document}

\title{Supersymmetric Large Extra Dimensions and the Cosmological Constant Problem}
\author[C.P. Burgess]{C.P. Burgess}
\address{Department of Physics and Astronomy, {\it McMaster University}, Hamilton ON
L8S 4M1, Canada and the {\it Perimeter Institute}, Waterloo ON,
N2L 2Y5, Canada. \email{{\sl cburgess@perimeterinstitute.ca}}}

\shortauthor{Burgess}

\maketitle
\begin{abstract}
This article briefly summarizes and reviews the motivations for
--- and the present status of --- the proposal that the small size
of the observed Dark Energy density can be understood in terms of
the dynamical relaxation of two large extra dimensions within a
supersymmetric higher-dimensional theory.
\\\\PACS Nos.:  31.15.Pf, 31.30.Jv, and 32.10.Hq
\end{abstract}
\begin{resume}
French version of abstract (supplied by CJP if necessary)
   \traduit
\end{resume}

\def\tablefootnote#1{%
\hbox to \textwidth{\hss\vbox{\hsize\captionwidth\footnotesize#1}\hss}}

\section{Reading the Tea Leaves}

The start of the new millenium finds the fundamental sciences at
an odd cross-roads. On one hand we have in the Standard Model
(which I take to include General Relativity and neutrino masses)
an exquisitely accurate description (at least in principle) of all
experiments that have ever been done. On the other hand we also
understand the limitations of this theory, which tells us it must
break down at sufficiently high energies. What might be the new
theory which replaces the Standard Model at higher energies, and
are we likely to discover it in the near future?

Much in our science depends on the answer to these questions, and
the nature of the answers depends crucially on precisely at which
energies the Standard Model fails. We know that the physics of the
longitudinal modes of the $W$ and $Z$ bosons --- {\it i.e.} the
physics of electro-weak symmetry breaking --- must lie just beyond
our present experimental reach (at a few TeV or less), and this is
the ultimate rationale behind the construction of the {\em Large
Hadron Collider} at CERN. But what happens beyond this depends on
more of the details, depending on what is found. Various clues
have been proposed over the years, motivated by understanding the
relative strength of the weak and gravitational interactions (the
electro-weak hierarchy); the nature of CP violation in the strong
interactions; the patterns of masses and mixings of the known
elementary particles; or the small observed gravitational response
to the energy of the vacuum.

All of the above clues share the feature that they demand an
explanation for one of the Standard Model's small dimensionless
numbers. The reason why such small numbers provide a useful clue
for the nature of new physics is based in how we understand small
numbers in modern quantum field theories. Suppose a parameter
$\lambda$ is found to be small when measured in an experiment
which is performed at an energy scale $\mu$. We wish to understand
this in terms of a microscopic theory of physics which is defined
at energies $\Lambda \gg \mu$, in terms of which the prediction
for $\lambda$ is given by
\begin{equation}
    \lambda(\mu) = \lambda(\Lambda) + \delta \lambda(\mu,\Lambda)
    \,.
\end{equation}
Here $\lambda(\Lambda)$ represents the direct contribution to
$\lambda$ due to the parameters in the microscopic theory, and
$\delta\lambda$ represents the contributions to $\lambda$ which
are obtained as we integrate out all of the physics in the energy
range $\mu < E < \Lambda$.

Science has encountered an enormous number of examples of small
parameters like this --- ranging from the ratio of the size of the
nucleus to the size of atoms in atomic physics, to the ratio of
the small superconducting gap to the typical electron energy in
condensed-matter physics. In {\em every single case} (so far) we
understand the smallness of $\lambda(\mu)$ in two steps: ($i$) the
quantity $\lambda(\Lambda)$ is understood to be small in the
microscopic theory, and ($ii$) the process of integrating out from
$\Lambda$ to $\mu$ gives an equally small contribution $\delta
\lambda$. In this way the smallness of $\lambda(\mu)$ can be
understood for any choice we may choose to make for the
microscopic scale $\Lambda$. The ubiquity of this kind of
explanation for small quantities has earned it a name: `{\em
technical naturalness}'.

Red flags are raised when both steps $(i)$ and $(ii)$ cannot be
understood, and a naturalness problem is declared for the
corresponding small quantity, $\lambda(\mu)$. The significance of
such naturalness problems comes from the clues they provide about
the existence of new physics. In particular, although we may not
be able to understand why $\lambda(\Lambda)$ is small until we
have the ultimate microscopic theory in our hands, we do expect to
be able to understand why the ordinary physics {\it which we
believe we already understand} at energies $\mu < E < \Lambda$
should not make $\delta \lambda(\mu,\Lambda)$ unacceptably large.
If we find $\delta\lambda$ to be many orders of magnitude larger
than the measured value $\lambda(\mu)$ then we suspect that we do
not actually understand the energies $\mu < E < \Lambda$ as well
as we thought - and progress can be made because this suggests
changes are required in an energy range which is experimentally
accessible.

The gravitational response of the vacuum energy is the case which
most egregiously resists such a technically-natural explanation
\cite{Weinberg}. In this case cosmological observations indicate
that the vacuum's energy density is at present at most $\rho \sim
(10^{-3}$ eV$)^4$ (in units with $\hbar = c = 1$)
\cite{ccnonzero}. But since a particle of mass $m$ typically
contributes an amount $\delta \rho \sim m^4$ when it is integrated
out, such a small value for $\rho$ can only be understood in a
technically natural way if $\Lambda \sim 10^{-3}$ eV or less.
Since essentially all of the elementary particles we know
(including the electron, for which $m_e \sim 5 \times 10^5$ eV)
have $m \gg 10^{-3}$ eV, it has proven impossible to understand
why our understanding of ordinary particles agrees with
experiments so well, and yet so badly predicts the vacuum's
observed gravitational response.

\section{SLED}

\subsection{Motivation}

A technically natural explanation for the vacuum energy density
requires two things. First, it requires a fundamental modification
of how gravity responds to physics at scales $E > \mu \sim
10^{-3}$ eV. Second, whatever provides this modification must not
ruin the excellent agreement with all of the many
non-gravitational experiments which have been performed over the
years for the energies $\mu < E < \Lambda$, with $\Lambda \sim
10^{11}$ eV, to which we have experimental access. It is the
mutual contradictions of these two conditions which has made it so
difficult to make progress.

Remarkably, a framework now exists within which both of these
conditions may be able to coexist: the framework of Large Extra
Dimensions \cite{ADD}. According to this picture --- which is
motivated by the discovery of $D$-branes within string theory ---
all of the observed particles apart from the graviton may be
trapped on a (3+1)-dimensional surface within an extra-dimensional
space. In such a world the presence of the extra dimensions would
only make a difference for gravitational observables, since only
gravitational probes could be used to probe for their existence.
Remarkably, the present upper limit for the size of such extra
dimensions is $r < 100 \; \mu$m, or $1/r > 10^{-3}$ eV
\cite{NewtonsLawTests} --- very close to the scale $\mu$ above
which our natural understanding of the vacuum energy breaks down!
Furthermore, extra dimensions can only be this large if there are
precisely two of them, and if the fundamental scale, $M_g$, of the
extra-dimensional physics is around 10 TeV, due to the relation
$M_p = M_g^2 \, r$ which relates $M_g$ and $r$ to the observed
Planck mass: $M_p = (8 \pi G)^{-1/2} \sim 10^{27}$ eV (where here
$G$ denotes Newton's constant). Other constraints from
astrophysics can also be satisfied for these choices of scales
\cite{PDG,LEDastrobounds,HR}.

{}From this we see that it is logically possible that the
gravitational response of the vacuum could depart dramatically
from our 4-dimensional expectations at precisely the scales $E >
10^{-3}$ eV where these expectations cannot account for the vacuum
energy in a technically natural way. This observation leads one to
ask what the gravitational response of the vacuum might be in such
a framework, and to ask whether the gravitational response of the
6D theory can be much smaller than would be expected from the 4D
perspective.

\subsection{6D Gravitational Response}

This line of thought leads to the proposal of Supersymmetric Large
Extra Dimensions (SLED), which posits there two large ({\it i.e.}
$r \sim 10$ $\mu$m) extra dimensions, arising within a
supersymmetric theory
\cite{Towards,susyADS,Update,MSLED,GGPplus,SLEDscalars,UVSensitivity,TAMU}.
The motivation for the large extra dimensions is as above ---
having extra dimensions this large provides a loophole for the
usual (four-dimensional) arguments which say that the vacuum
energy must be of order $m^4$, and so be too large. Supersymmetry
is motivated partially because our best high-energy theories are
supersymmetric, and partially because of the cancellations between
bosons and fermions which appears in the vacuum energy. This
proposal is briefly summarized here, following the discussion of
ref.~\cite{TAMU}.

In the SLED picture supersymmetry must be badly broken on our
brane, since we know that there are no super-partners for the
observed particles having masses which are much smaller than
$M_g$. Given this scale for supersymmetry breaking on the brane
there is also a trickle-down of supersymmetry breaking to the
`bulk' between the branes, whose size is set by the bulk's
Kaluza-Klein scale, $m_{sb} \sim M_{\rm KK}$, and so which for
unwarped geometries can be as low as $m_{sb} \sim 1/r \sim
10^{-2}$ eV \cite{susyaddbounds,Towards}. Much of the success of
the SLED proposal relies crucially on the ability to maintain this
hierarchy between the scales of supersymmetry breaking on the
brane and in the bulk (for other approaches to separating the
supersymmetry breaking scale see ref.~\cite{sepsusyscales}).

Within the above framework gravitational physics is effectively
6-dimensional for any energies above the scale, $1/r \sim 10^{-2}$
eV, and so the cosmological constant problem must be posed within
this new context. In order to see how the cosmological constant
problem is phrased in 6 dimensions, one must integrate out the
degrees of freedom between the scales $M_g \sim 10$ TeV and $1/r
\sim 10^{-2}$ eV. We seek the cosmological constant within the
effective 4D theory obtained after performing this integration,
which describes gravitational physics (like present-day cosmology)
on scales much larger than $r$. Imagine, therefore, performing the
integration over modes having energies $1/r < E < M_g$ in the
following three steps \cite{Towards}:
\begin{enumerate}
\item First, integrate out (exactly) all of the degrees of freedom
on the branes, to obtain the low-energy brane dependence on the
massless 4D graviton mode. In so doing we obtain (among other
things) a large effective brane tension, $T \sim M_g^4$ for each
of the 3-branes which might be present, which includes the vacuum
energies of all of the presently-observed elementary particles.
\item Next, perform the classical part of the integration over the
bulk degrees of freedom. This amounts to solving the classical
supergravity equations to determine how the extra dimensions curve
in response to the brane sources which are scattered throughout
the extra dimensions. It will be argued that it is this classical
response which cancels the contributions from the branes obtained
in Step 1, above.
\item Finally, perform the quantum part of the integration over
the bulk degrees of freedom. Given the cancellation of the
previous two contributions, it is this contribution which is
responsible for the fact that the present-day Dark Energy density
is nonzero. It is argued below that in some circumstances this
quantum contribution is of order $m_{sb}^4$, where $m_{sb} \sim
M_g^2/M_p \sim 10^{-2}$ eV is the supersymmetry-breaking scale in
the bulk. The small size of the 4D vacuum energy is in this way
attributed to the very small size with which supersymmetry breaks
in the bulk relative to the scale with which it breaks on the
branes.
\end{enumerate}

In a 4D world, the only contribution we would have is that of Step
1, above, and the problem is that this is much too large. But in a
6D world, because all of the observed particles are localized on
our brane their vacuum energy should be thought of as a localized
energy source in the extra dimensions, to which the bulk geometry
must respond. We next argue that the classical part of this bulk
response (Step 2, above) is of the same order as the direct
contribution of Step 1, and precisely cancels it in a way which
does not depend on the details of the supergravity involved or of
the precise extra-dimensional geometry which lies between the
various branes. In this way it provides a 6 dimensional
realization of self-tuning, whereby the effective 4D cosmological
constant is automatically adjusted to zero by the classical
response of the 2D bulk to the brane sources. The final nonzero
result finally comes from Step 3, due to the quantum bulk
contributions. But because the supersymmetry breaking scale in the
bulk is so small, these bulk loops are arguably the proper size to
agree with the recently-observed Dark Energy density. (It is this
part of the argument which non-supersymmetric proposals crucially
miss \cite{Precursors}.)

Step 1 consists of exactly integrating over all brane fields
having masses larger than $1/r$, and this produces a variety of
local interactions in the effective theory for energies $E \lsim
1/r$ on the brane. Since our interest is in the dependence of the
effective theory on the 4D metric, and we assume a large volume
for the extra 2 dimensions -- $M_g r \gg 1$ -- we may expand these
effective interactions in powers of the curvature:
\begin{equation}
    {\cal L}_b = - \sqrt{-g} \, \left[ T_b +
    \frac12 \, \mu_b^2 \, R + \cdots \right]\,,
\end{equation}
where on dimensional grounds we expect $T_b \sim M_g^4$, $\mu_b^2
\sim M_g^2$ {\it etc}..

Step 2 consists of the classical part of the bulk integration, and
so is equivalent to substituting into the classical action the
bulk field configurations which are found by solving the classical
field equations using the above effective brane action as a
source. It happens that for many situations this classical
response of the extra dimensions can be computed explicitly within
the approximation that the branes are regarded as delta-function
tension sources.

In the absence of nontrivial brane couplings to bulk fields like
the dilaton or bulk gauge fields (this assumption is relaxed
below), co-dimension 2 objects the extra-dimensional curvature
tensor typically acquire a delta-function singularity at the
position of the branes, corresponding geometrically to the
presence of a conical defect at the brane position. Einstein's
equations require that the singular contribution to the
two-dimensional curvature is given by
\begin{equation} \label{Ricci2}
    \sqrt{g_2} \, R_2 = - 2\, \sum_b T_b \,
    \delta^2(y-y_b) + \hbox{(smooth contributions)} \, ,
\end{equation}
Here $y_b$ denotes the position of the `$i$'th brane in the
transverse 2 dimensions, and the `smooth contributions' are all of
those which do not involve a delta-function at the brane
positions.

The effective 4D cosmological constant obtained after performing
Steps 1 and 2 above is obtained by plugging the above expression
into the classical bulk action. The effective 4D cosmological
constant obtained at this order is then
\begin{equation} \label{rhocl}
    \rho_{\rm cl} =  \sum_b T_b + \int_M d^2y \; e_2 \,
    \left[\frac12 \, R_2 + \dots \right]
    = 0 \, ,
\end{equation}
where the sum on `$b$' is over the various branes in the two extra
dimensions and `$\dots$' denotes all of the other terms besides
the Einstein-Hilbert term in the supersymmetric bulk action. The
final equality here has two parts. First, the sum over brane
tensions, $T_b$, precisely cancels the contribution of the
singular part of the curvature, eq.~(\ref{Ricci2}), to which they
give rise \cite{CLP}. Second, for supersymmetric theories a
similar cancellation also occurs amongst the various `smooth'
contributions in $\rho_{\rm cl}$ once these are evaluated for all
of the bulk fields using the classical field equations
\cite{Towards}. Interestingly, this cancellation does not depend
on the details of the bulk geometry, or on the number of branes,
since it relies only on a classical scale invariance which all 6D
supergravity actions enjoy \cite{susyADS}. Best of all, this
cancellation does not depend at all on the {\it value} of the
brane tension, $T_b$, and so applies equally well even if these
tensions are large and include all of the quantum effects due to
virtual particles localized on the branes.

We are left with the contribution of quantum effects in the bulk
(Step 3), to which we return in more detail in subsequent
sections. These must ruin the brane-bulk cancellation because the
scale invariance of the classical supergravity equations is not a
bona-fide quantum symmetry. However the bulk sector of the theory
is also one which is almost supersymmetric, since the bulk
supersymmetry-breaking scale is very small: $m_{sb} \sim 1/r \sim
10^{-2}$ eV. As a result we might expect standard supersymmetric
cancellations to suppress the quantum part of the result by powers
of $m_{sb}^2$, and if the leading term should be of order
$m_{sb}^4$ this would be the right size to account for the
observed Dark Energy density.

Under certain circumstances this is indeed what happens for some
6D supergravities \cite{UVSensitivity}. Quantum corrections do
lift the flat directions of the classical approximation, and those
loops involving bulk fields do so by an amount which is of order
$V(r) \sim m_{sb}^4$, leading to
\begin{equation} \label{eq:Vofr}
    V(r) \sim \frac{1}{r^4} \Bigl( a + b \log r \Bigr) \,,
\end{equation}
where $a$ and $b$ are calculable constants and the logarithmic
corrections generically arise due to the renormalization of UV
divergences in even dimensions \cite{casimirspheres}. The result
is this small despite the fact that the bulk loops include an sum
over Kaluza-Klein modes having 4D masses right up to the TeV
scale, $M_g$, due to cancellations which the extra-dimensional
supersymmetry enforces.

What is interesting about the potential, eq.~(\ref{eq:Vofr}), is
that it falls into a category of potentials which can provide a
phenomenologically viable description of the Dark Energy
\cite{CS}. Furthermore, this remains true even though there are
additional constraints which arise due to the extra-dimensional
interpretation of the Dark Energy. In particular, although the
cosmological evolution of the extra-dimensional volume can imply a
potentially dangerous time-dependence of Newton's constant over
cosmological epochs \cite{Nilles}, the existing bounds which
constrain how much this can happen are fairly easily satisfied due
to the effects of Hubble friction during the Universe's expansion
\cite{DEViable}. Furthermore, this potential predicts that
scalar-potential domination occurs when $\log (M_p r)$ is of order
$a/b$, which can easily be the required value given a modest
hierarchy amongst the coefficients, $a/b \sim 70$.

\subsection{The More General Case}

The above arguments assume particularly simple couplings between
the branes and the various bulk fields like the dilaton, and these
assumptions appear to play an important role in the
scale-invariance properties which underly the cancellation between
brane and bulk contributions \cite{susyADS}. It is therefore
natural to wonder what happens if these assumptions are relaxed. A
good test of how these arguments generalize is based on a class of
solutions to 6D chiral, gauged supergravity obtained in
ref.~\cite{GGP} by Gibbons, Guvens and Pope (henceforth GGP). What
makes these solutions so useful as a test of self-tuning is that
these authors derive the {\it most general} solution to these
field equations subject to two assumptions: ($i$) maximal symmetry
in 4 dimensions ({\it i.e.} de Sitter, Minkowski or anti-de Sitter
space), and ($ii$) axial symmetry in the internal 2 dimensions.
That is, they find the most general solutions whose metric has the
form
\begin{equation}  \label{GGPmetric}
    \exd s^2 =  W^2(\theta) g_{\mu\nu}(x) \, \exd x^\mu \exd
    x^\nu + \exd\theta^2 + a^2(\theta) \exd \varphi^2 \,,
\end{equation}
and for which $\phi = \phi(\theta)$ and $A_\phi = A_\phi(\theta)$.
Here the intrinsic 4D metric, $g_{\mu\nu}$, satisfies
$R_{\mu\nu\lambda\rho} = c (g_{\mu\lambda} g_{\nu\rho} -
g_{\mu\rho} g_{\nu\lambda})$, for some constant $c$. The assumed
axial symmetry corresponds to shifts of the coordinate $\varphi$,
and the metric can have singularities at up to two positions,
$\theta = \theta_\pm$, within the internal 2 dimensions
corresponding to the positions of source branes, but these
singularities are not required to be only conical in form. GGP
find that there is a five-parameter family of solutions to the
supergravity equations subject to these symmetry conditions. What
is most remarkable about these solutions is that {\it every single
one of them has a flat intrinsic 4D geometry} ({\it i.e.} $c =
0$), even though none of them is supersymmetric (except for a
single solution containing no branes), as is consistent with the
expectation that the effective 4D cosmological constant vanishes.
This same intrinsic flatness also appears to apply to the known
solutions which lie outside of the GGP assumptions
\cite{NewSolutions}.

Having the most general solutions, even subject to a symmetry
ansatz, also allows some exploration of how generic is the
classical cancellation of the effective 4D cosmological constant.
In order to do so it is useful to keep track of the physical
meaning of the parameters on which the general GGP solutions
depend. There are 5 such parameters, but one of these simply
parameterizes the flat direction whose existence is guaranteed by
the classical scale invariance of the supergravity equations. A
second parameter corresponds to another classical scaling
property, under which a redefinition of the fields may be used to
rescale the gauge coupling $g$ to any positive fixed value. The
three remaining parameters are broadly related to the three
physical quantities which characterize these geometries: the
tensions, $T_\pm$, of the two branes which source the bulk
geometry; and the overall magnetic flux of the background magnetic
field which (marginally) stabilize it.

\subsubsection{Topological Constraints}

We now ask what may be said about the natural of self-tuning given
the properties of these general solutions. There is a non-trivial
constraint amongst the parameters of the model which hold quite
generally for all of the GGP solutions, and it is natural to think
that these constraints hide the fine-tunings which underlie the
flatness of the 4D geometries. In fact, they do not because they
have their origins in topology, as is now explained.

There are two topological conditions which the GGP solutions all
share: one which expresses that the internal 2D geometry is
topologically a sphere; and one which expresses the quantization
(and so also conservation) of magnetic monopole flux
\cite{Towards,GibbonsPope}. Since the first of these turns out to
hold for all values of the parameters describing the classical
solution, it is of less interest as a potential source of
fine-tuning. It is the quantization of monopole number which
directly imposes a relation between the brane tensions, the gauge
couplings and one of the 5 parameters characterizing the various
GGP solutions.

The resulting topological constraint can be written in the
following way \cite{GGPplus}:
\beq \label{tensiontopology}
    \frac{g^2 e^{-\phi_0/2}}{2} \left( \frac{T_+ - T_-}{4 \pi}
    \right) = N^2 \left( \frac{g^2}{\tilde{g}^2} \right) \,,
\eeq
where $N$ is the integer which labels the monopole number. Here
$T_\pm$ are the two brane tensions, $g$ is the gauge coupling
which appears explicitly in the 6D supergravity action, $\tilde g$
is the gauge coupling for the background magnetic field and
$\phi_0$ is an additive constant in the dilaton configuration (and
so is one of the parameters describing the solution).

Although this looks like a hidden fine-tuning, first impressions
deceive \cite{susyADS,Update}. Recall in this regard that the
crucial issue for fine-tuning is whether or not the constraint is
stable against renormalization. That is, if
eq.~(\ref{tensiontopology}) is imposed amongst the renormalized
quantities at the TeV scale, does it automatically remain imposed
as successive scales are integrated out down to the scales below 1
eV? If so, then the constraint is technically natural, in the
sense described above, and so is not fine-tuned this (the most
serious) notion of fine tuning. But topological constraints are
{\it always} natural in this sense, because the integrating out of
successive scales of physics is a continuous process, and since
topological constraints involve quantization of quantities in
terms of integers, they remain unchanged by any such continuous
process. Topological constraints express global integrability
conditions which must be satisfied in order for solutions to
exist, rather than relations which select out a special class of
(flat) solutions amongst a wider class which do not have this
property.

\subsubsection{Runaway Solutions}

However one thing does emerge from an analysis of the properties
of the general solutions \cite{GGPplus} is the observation that
not all initial brane configurations can give rise to classically
stationary solutions. To see this notice that it turns out that
only a subset of the general GGP solutions involve purely conical
singularities, with the subset defined by the family of GGP
solutions whose tensions satisfy the condition
\cite{susyADS,GGPplus}:
\beq \label{Tconstraint}
     \left(1 -
    \frac{T_+}{4\pi} \right) \left( 1 - \frac{T_-}{4\pi} \right)
    = \frac{g^2\, e^{-\phi_0/2}}{2} \left( \frac{T_+ - T_-}{4 \pi}
    \right) = N^2 \left( \frac{g^2}{\tilde{g}^2} \right) \,.
\eeq
Here the last equality follows from using the topological
constraint, eq.~(\ref{tensiontopology}). Purely conical
singularities are only possible for a one-parameter locus of
tensions within the $T_+ - T_-$ plane.

Since brane solutions have conical singularities only in the
absence of couplings to the dilaton (in the Einstein frame)
\cite{AndrewT} it should be possible to choose a configuration of
branes whose tensions do not satisfy eq.~(\ref{Tconstraint}), and
so no bulk solution satisfying the GGP assumptions can exist for
such a choice. One of the assumptions must fail for such a
configuration and this is most likely the assumption of a static
bulk geometry, most likely leading to a runaway configuration.
Such a situation would be similar to what obtains if an arbitrary
configuration of electric charges were assembled: generically the
forces between them do not balance and so they move to zero or
infinite separation.

However the existence of a runaway need not in itself be a
problem. After all, we have seen that even if the classical
solution is static, a runaway is generated by quantum corrections.
Furthermore, this runaway can describe the Dark Energy density
provided $V \sim 1/r^4$, up to logarithmic corrections. If we
suppose that such classical runaways exist, then the central
question is: {\sl Is the classical runaway too steep to describe
the Dark Energy?} This need not be a problem even if so, since we
need not demand that {\it all} solutions describe our world, and
we know that {\it some} solutions do not admit classical runaways.
In this case we must know if it is technically natural to demand
that we find ourselves in a classically-static configuration.

A second question of central importance is: {\sl Are there hidden
self-tunings, in particular amongst the brane couplings (for which
supersymmetry cannot come to the rescue)?} A proper understanding
of this requires a detailed understanding of the matching
conditions between brane properties and bulk solutions, and
although the answer to this question is not yet clear it is at
present under active study \cite{Active,AndrewT}.

\section{Summary and Observational Consequences}

The SLED proposal states that the world becomes six-dimensional at
sub-eV energies, in such a way that the bulk gravitational physics
is supersymmetric down to the sub-eV KK scale. The main motivation
for this framework is that it dramatically changes the
gravitational response of the energy of the vacuum in a way which
appears to be technical natural, at least within the limits with
which it has so far been checked
\cite{Towards,susyADS,Update,MSLED,GGPplus}.

Regardless as to how the naturalness arguments turn out, the SLED
proposal dramatically changes how physics works at
experimentally-accessible energies, and so it is falsifiable
through the host of other robust phenomenological implications it
makes beyond those it has for cosmology. These include:
\begin{itemize}
\item Deviations from the inverse square law for gravity, which
more precise estimates show should arise for distances of order
$r/2 \pi \sim 1$ $\mu$m \cite{Callin};
\item A particular scalar-tensor theory of gravity at large
distances, with the scalar(s) being the moduli (like the volume)
which describe the two large extra dimensions. This is the same
scalar whose time-dependence now describes the Dark Energy
\cite{DEViable}.
\item Distinctive missing-energy signals in collider experiments
at the LHC due to the emission of particles into the extra
dimensions \cite{susyaddbounds,SLEDscalars}.
\item Potential astrophysical signals (and bounds) due to the
possibility of having too much energy loss into the extra
dimensions by stars and supernovae
\cite{PDG,LEDastrobounds,HR,susyaddbounds,nusled}.
\end{itemize}
If the SLED proposal is correct, it will be spectacularly so since
it requires this entire suite of observational implications to be
found. Indeed, it is this unprecedented connection between
observables in cosmology and particle physics --- which is driven
by its addressing the fundamental naturalness issues described in
previous sections --- that sets the SLED proposal apart from other
descriptions of Dark Energy.

\section*{Acknowledgements}
The SLED proposal represents the combined efforts of a number of
collaborators to whom I am most indebted, including Yashar
Aghababaie, Andy Albrecht, Georges Azuelos, Hugo Beauchemin,
Petter Callin, Jim Cline, Dumitru Ghilencea, Doug Hoover, Quim
Matias, Susha Parameswaran, Fernando Quevedo, Finn Ravndal,
Constantinos Skordis, Gianmassimo Tasinato, Andrew Tolley and
Yvonne Zavala. My funding during its development has come from
NSERC (Canada), FQRNT (Qu\'ebec), the Killam Foundation as well as
McGill and McMaster Universities. Many thanks to the organizers of
Theory Canada I for their kind opportunity to present these
results.

\end{document}